\documentclass[aps,onecolumn]{revtex4-2}
\usepackage[utf8]{inputenc}
\usepackage{graphicx}
\usepackage{xcolor}
\graphicspath{ {./img/} }

\begin{document}

\title{The Role of Low-energy ($<$ 20 eV) Secondary Electrons in the Extraterrestrial Synthesis of Prebiotic Molecules}

\author{Qin Tong Wu}
\affiliation
{Department of Chemistry, Wellesley College, Wellesley, MA 02481, USA}

\author{Hannah Anderson}
\affiliation
{Department of Chemistry, Wellesley College, Wellesley, MA 02481, USA}

\author{Aurland K. Watkins}
\affiliation
{Department of Chemistry, Wellesley College, Wellesley, MA 02481, USA}

\author{Devyani Arora}
\affiliation
{Department of Chemistry, Wellesley College, Wellesley, MA 02481, USA}

\author{Kennedy Barnes}
\affiliation
{Department of Chemistry, Wellesley College, Wellesley, MA 02481, USA}

\author{Marco Padovani}
\affiliation{INAF – Osservatorio Astrofisico di Arcetri, Largo E. Fermi, 5, 50125 Firenze, Italy}

\author{Christopher N. Shingledecker}
\affiliation{Department of Physics \& Astronomy, Benedictine College, Atchison, KS 66002, USA}

\author{Christopher R. Arumainayagam}
\affiliation
{Department of Chemistry, Wellesley College, Wellesley, MA 02481, USA}

\author{James B. R. Battat}
\email{jbattat@wellesley.edu}
\affiliation{Department of Physics \& Astronomy, Wellesley College, Wellesley, MA 02481, USA}


\begin{abstract}
We demonstrate for the first time that Galactic cosmic rays with energies as high as $\sim10^{10}$\,eV can trigger a cascade of low-energy ($<20$\,eV) secondary electrons that could be a significant contributor to the interstellar synthesis of prebiotic molecules whose delivery by comets, meteorites, and interplanetary dust particles may have kick-started life on Earth. For the energetic processing of interstellar ice mantles inside dark, dense molecular clouds, we explore the relative importance of low-energy ($<20$\,eV) secondary electrons—agents of radiation chemistry—and low-energy ($<10$\,eV), non-ionizing photons—instigators of photochemistry. Our calculations indicate fluxes of $\sim 10^{2}$\,electrons\,cm$^{-2}$\,s$^{-1}$ for low-energy secondary electrons produced within interstellar ices due to incident attenuated Galactic cosmic-ray (CR) protons. Consequently, in certain star-forming regions where internal high-energy radiation sources produce ionization rates that are observed to be a thousand times greater than the typical interstellar Galactic ionization rate, the flux of low-energy secondary electrons should far exceed that of non-ionizing photons. Because reaction cross-sections can be several orders of magnitude larger for electrons than for photons, even in the absence of such enhancement our calculations indicate that secondary low-energy ($<20$\,eV) electrons are at least as significant as low-energy ($<10$\,eV) non-ionizing photons in the interstellar synthesis of prebiotic molecules. Most importantly, our results demonstrate the pressing need for explicitly incorporating low-energy electrons in current and future astrochemical simulations of cosmic ices. Such models are critically important for interpreting James Webb Space Telescope infrared measurements, which are currently being used to probe the origins of life by studying complex organic molecules found in ices near star-forming regions. 
\end{abstract}

\maketitle

\section{Introduction}

Results of numerous experimental studies provide unambiguous evidence for the low-energy ($<$ 20 eV) electron-induced synthesis in interstellar ice analogs of complex organic molecules such as ethylene glycol \cite{harris_electron-induced_1995, schmidt_mechanisms_2022} and prebiotic molecules such as glycine, the simplest amino acid  \cite{Esmaili2018}. These experiments simulate submicron-size ice mantles surrounding carbonaceous or siliceous dust grains found within interstellar dark, dense molecular clouds, the birthplace of stars. In addition to condensed water, these cosmic ices are composed of ammonia, methanol, carbon dioxide, and other small molecules \cite{arumainayagam_extraterrestrial_2019}. These interstellar ice mantles, at temperatures as low as 10\,K, are bombarded by Galactic cosmic rays (CRs) which are composed of charged, high-energy particles (e.g., protons, electrons, and helium nuclei) that result from various mechanisms such as particle accelerations during supernova explosions, plasma shocks or stellar wind collisions \cite{cronin_cosmic_1997}. The energetic processing of interstellar ice mantles by high-energy CR particles and ionizing photons (e.g., Vacuum UV, X-rays, and $\gamma$-rays) is thought to be one of the mechanisms that initiate the extraterrestrial synthesis of prebiotic molecules such as cyanomethanimine (NC\textsubscript{2}HNH), which is a precursor of adenine, one of the four DNA nucleobases \cite{arumainayagam_extraterrestrial_2019}. In the early stages of our solar system, comets, asteroids, and meteorites carrying these prebiotic molecules may have delivered them to Earth, a likely critical step in the origin of life \cite{Altwegg2016}. The 2022 detection of (1) all the DNA/RNA nucleobases in carbonaceous meteorites \cite{oba_identifying_2022} and (2) several molecular precursors of RNA in a molecular cloud close to the center of the Milky Way \cite{rivilla_molecular_2022} provide tantalizing evidence for this posited mechanism which is now commonly termed molecular panspermia. 

While non-energetic processing (e.g., thermal chemistry \cite{theule_thermal_2013} and atom addition reactions \cite{linnartz_atom_2015}) may contribute significantly to the synthesis of prebiotic molecules, our research question involves determining the relative importance of two interstellar ice energetic processing mechanisms: photochemistry and radiation chemistry \cite{arumainayagam_extraterrestrial_2019}.
\begin{figure}[t]
\includegraphics[width=0.8\textwidth]{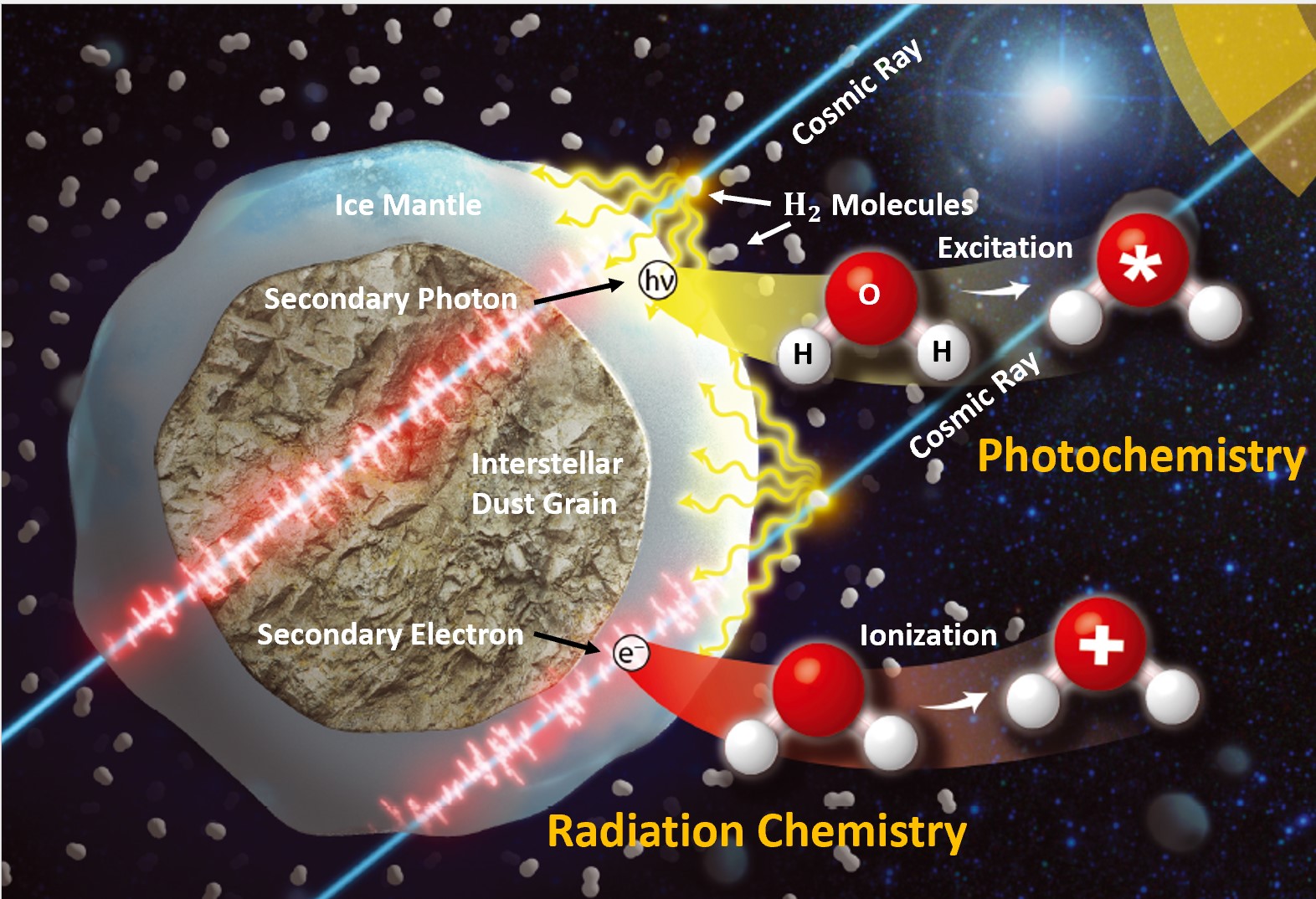} 
\caption{\label{fig:schematic}Schematic diagram demonstrating interstellar ices within dark clouds made of molecular hydrogen being processed by photochemistry involving the production of electronically excited water and radiation chemistry involving the production of ionized water and secondary electrons.
  This work focuses on the interaction depicted in the second of the two CRs (blue lines) in which a CR interacts in the ice mantle to produce secondary electrons.
  Reproduced from Ref. \citenum{arumainayagam_extraterrestrial_2019}. Copyright 2019, The Royal Society of Chemistry.}
\end{figure}
Photochemistry involves chemical processes that occur from the electronically excited state formed by photon absorption; during photochemistry, molecules absorb photon energy but are not ionized \cite{wayne_principles_1988,arumainayagam_photochemistry_2020}. Radiation chemistry involves chemical changes produced by the absorption of sufficiently high-energy (typically above 10 eV) radiation to produce ionization \cite{cooper_environmental_1998,arumainayagam_radiation_2020}. Low-energy secondary electrons, the production of which is a signature characteristic of radiation chemistry, are thought to be the dominant species in condensed-phase radiation chemistry \cite{Pimblott2007}. The role of photochemistry and radiation chemistry in the energetic processing of interstellar ices is depicted in Fig.~\ref{fig:schematic}. 

Due to low temperatures characteristic of star-forming regions, most chemical reactions are under kinetic control: reactions with low activation energies are favored. Accordingly, instead of the thermodynamic equilibrium constant, we consider the photolytic/radiolytic dissociation reaction rate constant $k$, which depends on the energy-dependent photon/electron flux $I(E)$ (the number of incident particles per unit time, unit energy, and unit area) and the dissociation reaction cross-section $\sigma(E)$ (also energy-dependent). As a first approximation, 
\begin{equation}
    k = \int I(E)\sigma(E)dE.
\end{equation}
Therefore, we attempt to answer our research question as follows. First, we compare the dissociation reaction cross-sections of low-energy electrons and photons. Second, we compare the flux of the Galactic-CR-induced low-energy secondary electrons (the driving force of radiation chemistry) produced \emph{within} the submicron-size ice mantles surrounding dust grains to the estimated flux of non-ionizing photons (the driving force of photochemistry) incident \emph{on} interstellar ice mantles.

Even for the simplest of molecules, no experimental data exists to compare directly condensed phase reaction cross-sections of low-energy photons and electrons as a function of incident energy. Nevertheless, theoretical considerations suggest that reaction cross-sections for electrons should be higher than those for photons \cite{Arumainayagam2010}. First, due to selection rules, governed primarily by dipole interactions and spin conservation, photon-molecule interactions are more restrictive compared to electron-molecule interactions. For example, unlike for electrons, photon-induced singlet-to-triplet transitions are nominally forbidden.
Second, in contrast to photons, electrons can be captured into resonant transient negative ion states which subsequently may dissociate \cite{Arumainayagam2010}. The resulting molecular fragments may then react with the parent molecule or other daughter products. Importantly, the total cross-section for dissociative electron attachment can be several orders of magnitude higher than the geometrical cross-section ($\sim 10^{-16}$\,cm$^{2}$) of a molecule.
For example, the total cross-section for dissociative attachment is $10^{-13}$\,cm$^2$ for producing Cl$^{-}$ from the electron attachment to CCl$_4$ \cite{Klar2001}.
Third, even though a typical molecule’s bond dissociation energy is relatively large ($\sim$ 5 eV), near-zero-energy electrons can cause a molecule to dissociate following electron attachment, especially for molecules that contain elements that have a high positive electron affinity, such as oxygen. Therefore, for incident energies below $\sim$ 5 eV, the probability for electron-induced dissociation is likely higher than for photon-induced dissociation because, for a given molecule, photons typically have a higher threshold energy for dissociation. 
Fourth, the electron attachment cross-section for non-polar molecules increases with decreasing electron energy for very low electron energies ($<0.3$\,meV) \cite{Fabrikant}.
Fifth, in contrast to photon impact electronic excitation, direct electron impact electronic excitation (not involving transient anions that decay into excited electronic states) is not an exclusively resonant process; the incident electron transfers that fraction of its energy sufficient to excite the molecule and any excess is removed by the scattered electron. Because direct electron-impact electronic excitation can operate over a wider range of energies than photon-impact excitation processes, direct electron-impact contribution to electronic excitation will be greater than any simple comparison of photon and electron-impact excitation cross-sections would suggest. Experimental results involving isolated CO molecules on amorphous solid water suggest that electrons are an order of magnitude more efficient than photons in promoting desorption \cite{Marchione2019}.
As a result of the many reasons enumerated above, reaction cross-sections are likely larger for electrons than for photons, particularly at incident energies corresponding to resonances associated with dissociative electron attachment.\footnote{Although reactive cross-sections for photons and electrons as a function of incident energy are unavailable, the photon absorption cross-sections and electron trapping cross-sections (the probability that an electron will be captured and localized) serve as upper limits to photodissociation and dissociative electron attachment cross-sections \cite{Bass1998}, respectively. For example, The photon absorption cross section for water has a maximum of $5\times 10^{-18}$\,cm$^2$ at $\sim 8.5$\,eV \cite{CruzDiaz2014}. In contrast, the electron trapping cross-section for condensed water varies from $\sim 3\times 10^{-16}$\,cm$^2$ near zero electron energy to $\sim 5 \times 10^{-17}$\,cm$^2$ at 10\,eV \cite{simpson}. The aforementioned limited experimental data suggest that reactive cross-sections for electrons are likely larger for electrons than photons, at least for water, the main constituent of interstellar and planetary ices.}

\begin{figure}[t]
\includegraphics[width=0.8\textwidth]{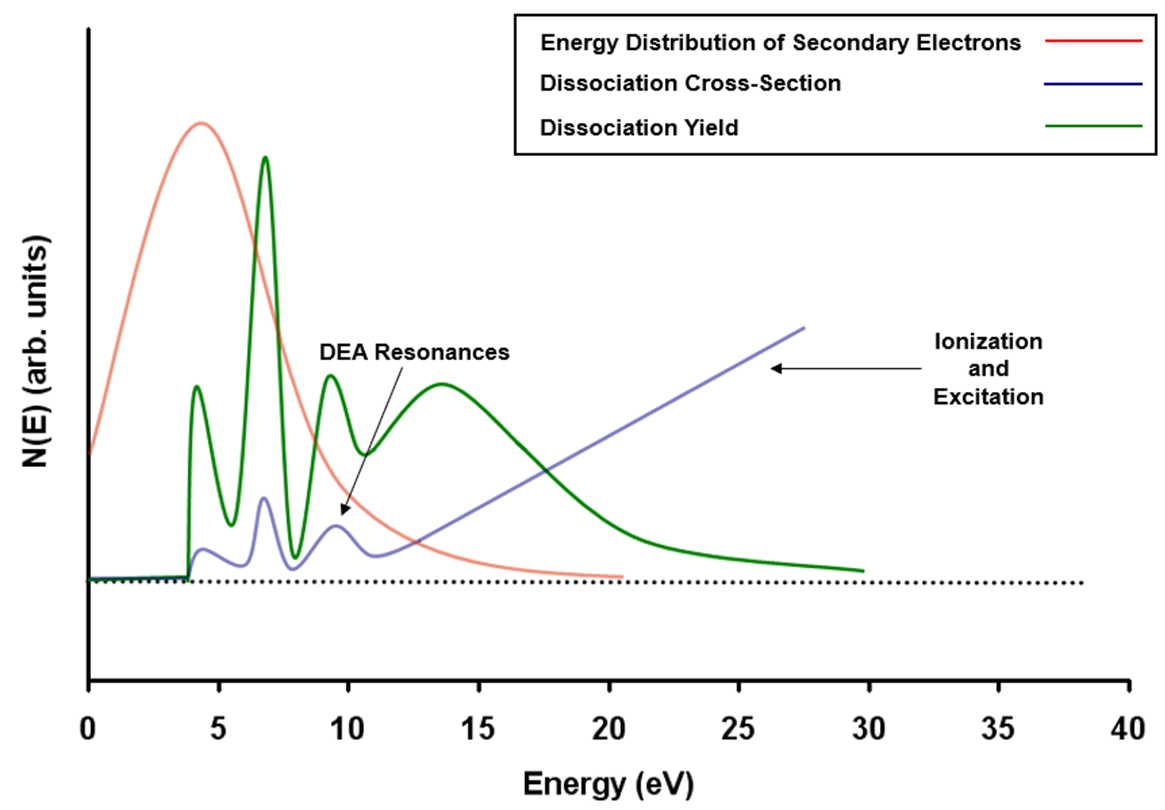} 
\caption{\label{fig:yield}Schematic of the energy distribution of secondary electrons generated during a primary ionizing event (red curve), 
    the cross-section for electron-induced dissociation for a typical molecule (blue curve), and the dissociation yield as a function of electron energy for a typical molecule (green curve). The green curve is the product of the red and blue curves. Reproduced from Ref. \citenum{Arumainayagam2010}. Copyright 2009, Elsevier B.V.}
\end{figure}

In addition to electrons having larger reaction cross-sections, electron-induced reactions may predominate over photon-induced reactions because of the sheer number of low-energy secondary electrons produced by high-energy irradiation. The interaction between high-energy radiation (e.g., $\gamma$-rays, X-rays, electrons, and ion beams) and matter produces copious numbers ($\sim 4 \times 10^{4}$ per MeV of energy deposited) of cations and non-thermal secondary low-energy electrons \cite{Kaplan}.
A significant majority of the incident radiation energy is transferred to the kinetic energy of secondary electrons \cite{Sanche2022}.
The inelastic collisions of these low-energy electrons with molecules and atoms produce distinct energetic species that are the primary driving forces in a wide variety of radiation-induced chemical reactions. Therefore, low-energy secondary electrons are thought to be the dominant species in condensed-phase radiation chemistry \cite{Pimblott2007}. Results of Monte Carlo simulations of high-energy radiation interacting with water demonstrate that nearly 90\% of the secondary and successive generations of secondary electrons have initial energies less than 20\,eV \cite{Cobut1998}. The most probable energy of the secondary electrons is $\sim 10$\,eV \cite{Pimblott2007}.
Even though the dissociation probability for a generic molecule increases monotonically with increasing incident electron energy from $\sim 10$ to 100\,eV due to dissociative electronic excitation and ionization, the dissociation yield is most significant at low ($< 20$\,eV) incident electron energies due to the abundance of secondary electrons at those energies.
Fig.~\ref{fig:yield} clearly demonstrates the importance of low-energy ($< 20$\,eV) secondary electrons in causing high-energy radiation-induced chemical changes \cite{Arumainayagam2010}.

Based on the arguments presented above, the primacy of low-energy electrons in the interactions of high-energy radiation with matter is now being exploited for practical applications in disparate fields, suggesting the universality of this phenomenon in radiation chemistry. One such application is cancer treatment \cite{Sanche2022, Rezaee}. Results of quantitative experimental and theoretical studies of low-energy electron-induced DNA lesions are being used to design targeted radionuclide therapy and nanoparticle-aided radiotherapy \cite{Sanche2022, SancheNano}. For example, recent studies suggest that radiation-induced bystander effects may be reduced by exploiting the low mean free path of low-energy electrons emitted by \textsuperscript{125}I-coated nanoparticles \cite{Sykes}. In addition to their applications in health care, low-energy electrons play an important role in many industrial processes involving ionizing radiation. Recent experimental and theoretical studies have also demonstrated the critical role that low-energy electrons play in Extreme ultraviolet (EUV) (92\,eV) lithography for the fabrication of the next generation sub-20\,nm scale semiconductor chips. For example, electrons with energies as low as 1.2\,eV can chemically modify EUV resists, as evidenced by a recent low-energy electron microscopy study \cite{bespalov}. It has long been realized that a fundamental understanding of the production and interactions of low-energy electrons is necessary for optimizing materials science techniques such as Focused Electron Beam–Induced Deposition \cite{de_vera_multiscale_2020} and plasma processing \cite{Mason2019}.

While fields such as cancer therapy and materials science have demonstrated tremendous progress in models that incorporate the role of low-energy electrons in the interactions of high-energy radiation with matter, current astrochemical models fail to explicitly account for low-energy secondary electrons produced within cosmic ices by incident high-energy radiation. One notable exception is a recent publication that partially accounts for the role of low-energy secondary electrons in the energetic processing of interstellar ices \cite{Shingledecker2019}. Interestingly, in 2018, glycine formation was observed in CO\textsubscript{2}:CH\textsubscript{4}:NH\textsubscript{3} ices irradiated by electrons with energies as low as 9\,eV \cite{Esmaili2018}. Low-energy electrons in the interstellar medium may result from two processes: (1) the interaction of Galactic CRs with the gaseous molecular hydrogen present in the dark, dense molecular clouds \cite{Irvine1991}, and (2) the inelastic collisions that ionizing radiation (e.g., Galactic CR) experiences as it traverses through the ice-covered dust grains \cite{Boamah2014}. Recent calculations indicate that an electron produced through process (1) strikes a dust grain in a dense molecular cloud once every 25,000 years \cite{Mason2019}, suggesting that low-energy electrons incident on ices play an insignificant role in interstellar chemistry. However, by focusing solely on incident electrons, these calculations ignore a more significant interstellar ice radiation chemistry driver: low-energy secondary electrons \textit{produced within} cosmic ices by Galactic CRs and high-energy radiation internal to the molecular cloud.
We estimate the flux of Galactic-CR-induced secondary electrons within interstellar ices by (1) considering the attenuated Galactic-CR particle spectra after propagation through dark, dense molecular clouds and (2) using data from the National Institute for Standards and Technology (NIST) PSTAR\footnote{https://physics.nist.gov/PhysRefData/Star/Text/PSTAR.html} 
database to account for the total stopping power (the energy loss per unit length) for protons in water (used as a model for ice, as discussed later). Our results indicate that the flux of Galactic-CR-induced-low-energy electrons \emph{within} interstellar ices is almost as substantial as that of non-ionizing UV photons incident \emph{on} ices in dark, dense molecular clouds. 

Attenuated Galactic CRs are not the only source of ionizing radiation incident on interstellar ices within dark, dense, molecular clouds. In certain stellar nurseries, the CR ionization rate of molecular hydrogen (hereafter CRIR) 
has been discovered to be a thousand times greater than typical observed Galactic values of 
$\sim10^{-17}$--$10^{-15}$~s$^{-1}$ \cite{fontani_seeds_2017, Ceccarelli_2014}. This enhancement in the CRIR has been attributed to embedded ionizing radiation sources within these star-forming regions. This high-energy radiation is likely due to populations of relativistic particles and their associated non-thermal synchrotron emission. 
Charged particles may be accelerated in protostellar jet shocks and in accretion shocks on protostellar surfaces.
Although protostellar jet velocities (tens to hundreds of km\,s$^{-1}$) are much smaller than relativistic speeds, these particles reach relativistic velocities through the Diffusive Shock Acceleration mechanism in which particles gain energy by diffusing back and forth across a shock or jet front \cite{1983RPPh...46..973D,Kirk1994,padovani_m_cosmic-ray_2015,padovani_m_protostars_2016}. Recent research has provided evidence to support this incipient theory.
For example, observations with the NRAO's Karl G. Jansky VLA of the Class I intermediate-mass protostar HOPS 370 of the Orion Molecular Cloud 2 located at a distance of $414\pm7$\,pc suggest that non-thermal synchrotron emission from relativistic electrons accelerated in shocks produce the observed non-thermal emission from knots (compact regions wihtin molecular clouds where gas and dust are concentrated).
Non-thermal synchrotron emission has also been detected in well-known protostellar jets, HH 80–81, located in the L291 cloud in Sagittarius at 1.7\,kpc, thus indicating that acceleration mechanisms exist within the jet and may be responsible for the enhanced CRIR \cite{rodriguez-kamenetzky_highly_2017}. Other deep radio continuum observations at 325 and 610\,MHz using the Giant Metrewave Radio Telescope of the young, low-mass star DG Tau located at 140\,pc in the Taurus Molecular Cloud provide tentative evidence for the acceleration of particles to relativistic energies due to the impact of a low-power jet suggesting that low-energy CRs are being generated by young, low mass stars \cite{ainsworth_tentative_2014}. Additionally, ALMA observations of the Class 0 protostar object B335 that is associated with an east-west outflow located at 164.5\,pc show very high CRIRs (between $10^{-16}$ and $10^{-14}$\,s$^{-1}$) that increase toward the central protostellar embryo, indicating that the local acceleration of CRs and not the penetration of interstellar Galactic CRs may be responsible for the gas ionization in the inner envelopes of the protostar \cite{cabedo_victoria_magnetically_2023}. Moreover, observations of FIR4, a young intermediate-mass protocluster of the Orion Molecular Cloud 2, using the IRAM NOEMA interferometer uncovered a possible jet shock propagating towards a previously measured enhanced CRIR region, suggesting that energetic particle acceleration by jets might be responsible for the enhanced CRIR in these regions \cite{lattanzi_v_solis_2023}. These recent observations suggest that internal high-energy ionizing radiation sources could be a dominant source of low-energy secondary electrons produced within ice mantles found inside dark, dense molecular clouds, the birthplace of molecules and stars.

Observations detailed above and our calculations described herein demonstrate that the flux of low-energy secondary electrons within interstellar ices produced by Galactic CRs and internal ionizing radiation may far surpass that of non-ionizing photons incident on interstellar ices. Because electrons likely have larger reaction cross-sections than photons, our calculations demonstrate the pressing need for astrochemical models to incorporate the role of low-energy ($<20$\,eV) electrons in the extraterrestrial synthesis of prebiotic molecules.

\section{Methods}

\subsection{Overview}
To compute the number of CR-induced low-energy electrons available for radiation chemistry in cosmic ices, we select a Galactic CR spectrum for protons and then compute how that spectrum is attenuated as the Galactic CRs traverse a molecular cloud composed primarily of molecular hydrogen. The protons in this attenuated Galactic spectrum subsequently interact with ice-covered dust grains, losing energy and producing secondary electrons that can contribute to radiation chemistry. In this section, we describe our choice of model for the initial Galactic CR proton spectrum, the procedure by which we compute the post-propagation spectrum and the methodology by which we estimate the number of Galactic-CR-induced electrons produced in the ice-covered interstellar dust grains.

\subsection{Galactic cosmic ray spectrum}

Galactic CR particles, with energies as high as $10^{20}$\,eV, consist of approximately 90\% protons, 9\% alpha particles, and 1\% heavier nuclei \cite{gaisser_cosmic_2016}. While the flux of hydrogen and helium nuclei dominate all other species, there is also a steady flux of CR electrons, positrons, and antiprotons \cite{stanev_high_2021}. The total Galactic CR-induced secondary electron flux is the sum of the secondary electron flux produced by CR protons, alpha particles, heavier nuclei, electrons, positrons, and antiprotons. Here, we restrict our calculations to Galactic CR protons.
We also ignore secondary electrons produced by ionizing Vacuum UV, $\gamma$-rays, and X-rays incident on ices within dark, dense molecular clouds. Most importantly, we do not take into account embedded ionizing radiation sources within star-forming regions inside molecular clouds. Therefore, our calculations represent a lower bound to the secondary-electron flux produced by ionizing radiation within interstellar ices, and yet we still find that the electron flux is significant.

The spectra of interstellar CR nuclei at high energy (above 1\,GeV\,nuc$^{-1}$) are well-constrained by ground, balloon, and satellite observations \cite{AMS:2014bun,AMS:2015tnn}; however, the low-energy nuclei are strongly influenced by solar modulation effects and their spectra are less well constrained \cite{Putze:2010fr,Indriolo2009}. Voyager 1 and 2 data \cite{Cummings:2016pdr,stone_cosmic_2019} provide the best constraints on interstellar CR spectra at low energies.

\begin{figure}
    \includegraphics[width=0.8\textwidth]{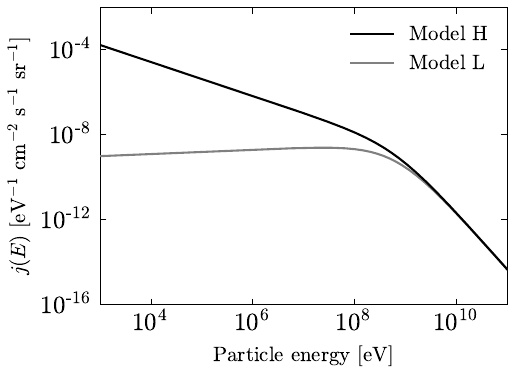} 
    \caption{\label{fig:inputSpectra}Differential interstellar CR spectra for protons  -- see Eqn.~(\ref{eq:specModel}) and Table~(\ref{tab:params}).} 
\end{figure}

In this work we use the analytical model for the interstellar CR spectrum of protons (and electrons) provided by Ivlev et al. \cite{Ivlev2015}:
\begin{equation}
\label{eq:specModel}
     j(E) =  C\frac{E^\alpha}{(E+E_{0})^\beta} \,\,\mbox{ eV}^{-1} \mbox{ cm}^{-2} \mbox{ s}^{-1} \mbox{ sr}^{-1}.
\end{equation}
We adopt the slightly modified parameter values for $E_0$ and $\alpha$, as advocated by Padovani et al. \cite{Padovani2018} 
The associated model parameter values are given in Table~\ref{tab:params}, and a plot of the spectra is shown in Fig.~\ref{fig:inputSpectra}. 

\begin{table}[t]
\begin{tabular}{l r r r r}
 \hline\hline
 \multicolumn{1}{c}{Species} & 
 \multicolumn{1}{c}{$C$} & 
 \multicolumn{1}{c}{$E_{0}$/MeV } & 
 \multicolumn{1}{c}{$\alpha$} & 
 \multicolumn{1}{c}{$\beta$} \\
 \hline
  $p$ (Model \textit{L}) &  $2.4 \times 10^{15}$ & 650 & 0.1 & 2.8\\
 \hline
  $p$ (Model \textit{H}) &  $2.4 \times 10^{15}$ & 650 & $-$0.8 & 1.9\\
 \hline
\end{tabular}
\caption{\label{tab:params} Parameters for the interstellar CR proton spectrum for Model L and Model H, as defined in Eq.~\ref{eq:specModel}.} 
\end{table}

We consider two models for the Galactic CR proton spectrum: Model L (``low") and Model H (``high"). Model L is based on Voyager 1 and Voyager 2 data collected within the Very Local Interstellar Medium \cite{stone_cosmic_2019}. The data were obtained when the spacecraft were at a heliocentric distance of 122 AU, just beyond the heliopause. Although Voyager 1 and Voyager 2  provide the only direct observational constraint currently available on the low-energy CR spectra, the measured proton flux is likely not the interstellar flux because the spacecraft had not yet entered interstellar space \cite{Ivlev2015}. Most importantly, Model L fails to reproduce the CRIR estimated from observations in diffuse clouds \cite{indriolo_investigating_2012}. Hence, Model H was introduced to ensure agreement with the average ionization rate of H$_2$, as derived from the measured abundance of H$_3^+$ in diffuse clouds \cite{Ivlev2015}. Models L and H are taken to be the lower and upper bounds on the proton Galactic CR spectrum. 

\subsection{Computing the local cosmic ray spectrum}
The interstellar CR spectrum is altered (attenuated) as it propagates through a molecular cloud. Molecular clouds consist primarily of H$_2$ (molecular hydrogen), with contributions from other species (He, C, N, O, etc.) typically less than 15\%. For this work, it is sufficient to approximate the cloud composition as 100\% H$_2$. Future work could consider more detailed interstellar medium compositions, such as that given in Wilms et al. \cite{2000ApJ...542..914W}. 

\begin{figure}[t]
\includegraphics[width=0.6\textwidth]{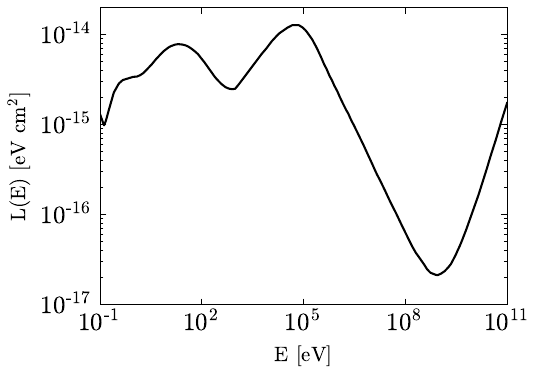}
\caption{\label{fig:energyLossProton}Energy loss function for protons interacting with H$_2$. Data from Ref.~\citenum{padovani_cosmic-ray_2009}.}
\end{figure}

To model the interactions of the Galactic CR protons with the molecular cloud, we adopt the \textit{continuous-slowing-down approximation} (CSDA). The CSDA assumes that the proton's direction of propagation does not change significantly from the interactions with the H$_2$ molecules, and that the energy loss function of the proton $L(E)$ is continuous along its path and proportional to $dE/d\ell$ the energy lost per unit path length $\ell$: 
\begin{equation}\label{eq:energyLoss}
L(E) = -\frac{1}{n(\mbox{H}_2)}\left(\frac{dE}{d\ell}\right),
\end{equation} 
where $n(\mbox{H}_2)$ is the number density of molecular hydrogen in the molecular cloud. 
We use the energy loss function for protons in H$_2$ as given by Padovani et al. \cite{padovani_cosmic-ray_2009}, and shown in Fig.~\ref{fig:energyLossProton}.  As explained below, the attenuated spectrum can be written analytically in terms of the Galactic spectrum $j(E)$ and the energy loss function.

The column density of molecular hydrogen $N(\mbox{H}_2)=\int n(\mbox{H}_2) d\ell$, can also be expressed in terms of the energy loss function using Eqn.~\ref{eq:energyLoss} as:
\begin{equation}\label{eq:colRange}
N(\mbox{H}_2) = -\int_{E_0}^E \frac{dE'}{L(E')} = n(\mbox{H}_2)\left[R(E_0)-R(E)\right].
\end{equation} 
We have introduced the range $R(E)$ of a proton of energy $E$, defined by:
\begin{equation}\label{eq:range}
R(E) = \frac{1}{n(\mbox{H}_2)}\int_0^E \frac{dE'}{L(E')},
\end{equation} 
and note that the energy of a proton decreases from $E_0$ to $E$ after interacting with a 
column density $N(\mbox{H}_2)$.

Our goal is to compute $j(E,N)$, the Galactic CR proton spectrum after traversing a column density $N$, in terms of the interstellar spectrum $j(E_0, 0)$ and the energy loss function $L(E)$ by
following the prescription of Takayanagi \cite{1973PASJ...25..327T} and further elaborated by Padovani et al. \cite{Padovani2018}. We assume that the number of CR protons is conserved, so that:
\begin{equation}\label{eq:consOfCRs}
    j(E,N)dE = j(E_0,0)dE_0.
\end{equation}
The total differential of Eqn.~\ref{eq:colRange} is given by  $dN(\mbox{H}_2) = \frac{\partial N(\textrm{\scriptsize H}_2)}{\partial E}dE + \frac{\partial N(\textrm{\scriptsize H}_2)}{\partial E_0} dE_0$. 

If we consider a fixed value of $N(\mbox{H}_2)$, so that $dN(\mbox{H}_2)=0$, then we find the following relation:\footnote{Notice that $N(\mbox{H}_2)$, as defined in Eqn.~\ref{eq:colRange}, is an explicit function of both $E_0$ and $E$ because those variables appear in the limits of the integral. By the Fundamental Theorem of Calculus, $\partial N/\partial E=-1/L(E)$, and similarly, $\partial N/\partial E_0 = +1/L(E_0)$.} 
\begin{equation}\label{eq:consOfCRs2}
    \frac{dE}{L(E)} = \frac{dE_0}{L(E_0)}. 
\end{equation}
Together, Eqns.~\ref{eq:consOfCRs} and \ref{eq:consOfCRs2} give our desired expression for the attenuated Galactic CR proton spectrum in terms of the Galactic spectrum at the nominal column density $N({\rm H_2}) = 0$, for a given value of $N(\mbox{H}_2)$:
\begin{equation}\label{eq:localSpec}
    j(E,N) = j(E_0,0)\frac{L(E_0)}{L(E)}.
\end{equation}
\begin{figure}[t]
\includegraphics[width=0.75\textwidth]{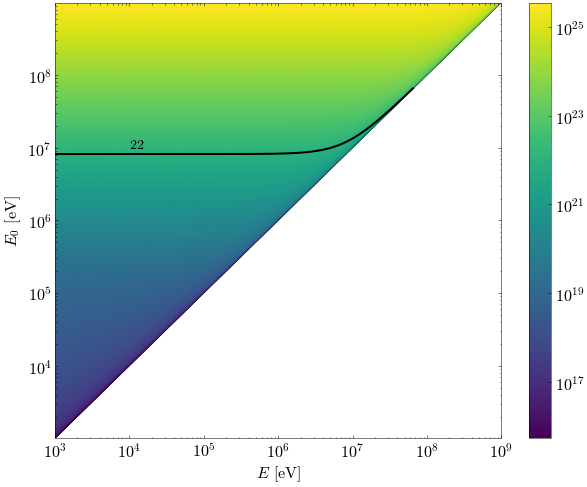}
\caption{\label{fig:contour}$N(\mbox{H}_2)$ values computed from Eqn.~\ref{eq:colRange} as a function of $E_0$ and $E$. The lower right region of the plot is empty because it corresponds to the unphysical situation $E>E_0$. Overlaid on the plot is a single contour line corresponding to the set of $E_0$ and $E$ values for which $N(\mbox{H}_2)=10^{22}$\,cm$^{-2}$ (so $dN=0$ along the contour).}
\end{figure}
All that remains is to determine, for a given $N(\mbox{H}_2)$, the final energy $E$ in the attenuated spectrum for each initial energy $E_0$ in the interstellar spectrum. For that, we first compute $N(\mbox{H}_2)$ for all values of $E$ and $E_0$, and then extract contours for a set of $N(\mbox{H}_2)$ values, as shown in Fig.~\ref{fig:contour} ($dN=0$ along a given contour). Each contour represents a mapping from $E_0$ to $E$ for a specific column density, which we then fit with the analytic function \cite{Padovani2018}
\begin{equation}\label{eq:E0vsE}
    E_0(E,N) = \left(cE^b+\frac{N}{N_0}\right)^{1/b},
\end{equation}
where $b$, $c$, and $N_0$ are fit parameters. Together, Eqns.~\ref{eq:localSpec} and \ref{eq:E0vsE} can be used to compute the attenuated Galactic CR spectrum $j(E,N)$ for a given column density $N(\mbox{H}_2)$.

\begin{figure}[t]
\includegraphics[width=\textwidth]{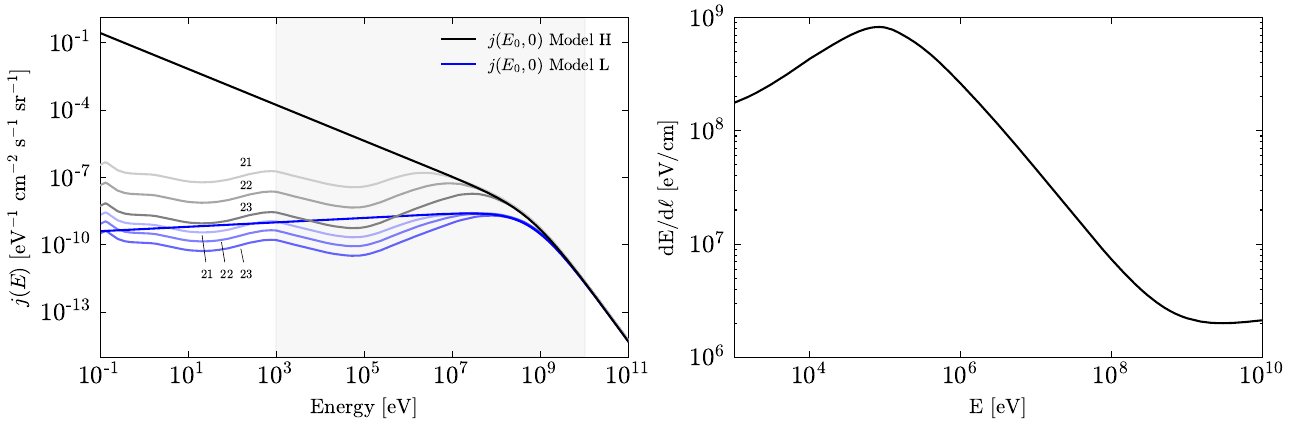}
\caption{\label{fig:inputAndDegradedSpectra}(Left): Interstellar CR proton spectra (black: Model H; blue: Model L) and attenuated Galactic CR spectra for three column densities $N(\mbox{H}_2)=10^{22\pm1}$\,cm$^{-2}$ (grayscale: Model H; blue-scale: Model L). Numbers on the plot indicate $\log_{10}[N(\mbox{H}_2)/\mbox{cm}^{-2}]$. (Right): Stopping power for protons in liquid water from the NIST PSTAR database \cite{Pstar}. We assume that the stopping power for protons in water ice is approximately the same as in liquid water. Data are presented as energy loss per path length traveled, assuming a water mass density of 1\,g\,cm$^{-3}$. Stopping data from NIST is only available for proton energies of 1\,keV to 10\,GeV, which corresponds to the lightly shaded region in the left-hand plot. }
\end{figure}

Figure~\ref{fig:inputAndDegradedSpectra} shows the Galactic proton spectra $j(E,0)$ (for Models H and L), along with the attenuated proton spectra $j(E,N)$ for column densities of $N(\mbox{H}_2) =10^{22\pm 1}$\,cm$^{-2}$, consistent with expectations for dark, dense molecular clouds.\footnote{Dense molecular cloud cores are usually defined as those regions where UV photons of the interstellar radiation field are absorbed so that only CRs penetrate the cloud. See e.g. Table 3.1 of Ref.~\citenum{2004fost.book.....S}.}

\subsection{Interactions of protons with cosmic ices}
Interstellar ices surrounding submicron-size dust grains exist with a broad range of sizes and compositions. Here we adopt the admittedly simplistic assumption of a sphere of diameter $D_{ice}=100$\,nm composed entirely of water ice even though interstellar ices contain in lower abundances species such as CO, CO$_{2}$, CH$_{4}$, and CH$_{3}$OH \cite{oberg2016}.  
We further assume that every proton striking the ice sphere will travel through its full diameter $D_{ice}$. To estimate the energy deposited in the ice by protons we need to know the stopping power of protons in the ice. Although the precise stopping power $dE/d\ell$ for protons in ice is not available, for our purposes it is adequate to use the stopping power of protons in \textit{liquid} water as a proxy. That data is available from the NIST PSTAR database \cite{Pstar}, and is shown in Fig.~\ref{fig:inputAndDegradedSpectra}. We further note that most protons in the local CR spectrum lose only a small fraction of their energy in the ice. 
For example, a 1\,MeV proton loses 3\,keV of energy over 100\,nm (less than 1\% of its energy). 
We, therefore, assign a constant stopping power to each proton as it travels through the ice. As shown in Fig.~\ref{fig:inputAndDegradedSpectra}, however, the stopping power has a strong energy dependence, which we account for in our analysis by assigning different stopping powers to protons of different initial energies.

We define the secondary electron flux $\Phi$ in the ice as the number of secondary electrons per area per time  produced in the ice by the attenuated Galactic CR protons:
\begin{equation}
    \Phi = \frac{2\pi D_{ice}}{w} \int_{E_{min}}^{E_{max}} j(E',N) \left.\frac{dE}{d\ell}\right|_{E'} \,dE'
\end{equation}
where $D_{ice}$ is the diameter of the ice grain (here, taken to be 100\, nm), the factor of $2\pi$ comes from integrating $j(E,N)$ over solid angle (and only considering protons incident on the exterior surface of the ice grain), and the stopping power $dE/d\ell$ in the integrand is evaluated at each energy of interest $E'$. The term $w$ is the \textit{differential} $w$-value which specifies the average energy required to make an electron-hole pair in the ice in the case where a small fraction of the energy of the proton is lost in the ice.\footnote{This is distinct from the integral $W$-value, which applies to particles that fully stop in the medium. In that case, nuclear interactions, which are important at low projectile energy, make $W\geq w$. In the limit of high projectile energy, $W$ approaches $w$.} We discuss the $w$-value more below. The integral is carried out over a range of proton energies $\left[E_{min},E_{max}\right]$. Our analysis is limited to energies for which stopping power data is available: $E_{min}=1$\,keV and $E_{max}=10$\,GeV. Ignoring protons with energy outside of this range will necessarily underestimate the number of secondary electrons produced in the ice. But as we will see, we still find that a significant number of secondary electrons are produced inside the ice by the incident attenuated Galactic CR protons.

We are not aware of any measurements or calculations of the $w$-value for protons in water ice; we make the reasonable assumption that the $w$-value for protons in liquid water is approximately the same as for protons in ice. 
Baek and Grosswendt \cite{Baek2007} calculate $w$ for protons in liquid water as a function of proton energy using three different models. All three provide similar results for large proton energies, with the $w$-values approaching 25--27\,eV. Here, we adopt a conservative value of 30\,eV.

\section{Results}

\begin{table}[t]
\centering
\begin{tabular}{||l r r||} 
 \hline
\multicolumn{1}{||c}{CR proton spectrum} & 
\multicolumn{1}{c}{$\log\left[N(\mbox{H}_2)/\mbox{cm}^{-2}\right]$} & 
\multicolumn{1}{c||}{$\Phi$} \\
\multicolumn{1}{||c}{} & 
\multicolumn{1}{c}{} & 
\multicolumn{1}{c||}{(e$^-$ cm$^{-2}$ s$^{-1}$) } \\
 \hline\hline
Model H & 21 &      500 \\
        & 22 &      200 \\
        & 23 &       90 \\
\hline
Model L & 21 &       20 \\
        & 22 &       20 \\
        & 23 &       10 \\
\hline
\end{tabular}
\caption{\label{tab:results}Calculated secondary electron flux $\Phi$ produced within interstellar ice mantles inside dark dense molecular clouds due to incident Galactic CR protons. Data are presented for Model H and L, and for three representative values of the column density $N(\mbox{H}_2)$. Values of $\Phi$ are rounded to one significant figure.}
\end{table}
Our estimates of the low-energy secondary electron flux produced by CR protons \emph{within} interstellar ices inside dark dense molecular clouds are shown in Table~\ref{tab:results}. For Model H, we find a flux of $2\times10^2$ electrons cm$^{-2}$\,s$^{-1}$ for a typical column density of $N(\mbox{H}_2) =10^{22}$\, cm$^{-2}$. For comparison, the accepted dense cloud photon flux incident \emph{on} interstellar ices is on the order of $10^3$ photons cm$^{-2}$\,s$^{-1}$ and only 60\% of the flux is estimated to be non-ionizing \cite{Prasad1983, arumainayagam_extraterrestrial_nodate}. Other factors such as the significantly smaller penetration depth of UV photons\footnote{In the condensed phase, 5–9\, eV photons, those likely responsible for most solid-state photochemistry, have mean free paths comparable to the thickness of interstellar ice. For example, the mean free path of 8.5\, eV photons in condensed water is $\sim$0.06 $\mu$m, according to calculations based on the photon absorption cross-section ($5\times10^{-18}$\,cm$^2$) of water ice. Because ice mantles surrounding cold interstellar grains are $\sim$ 0.1 $\mu$m, the photon flux will diminish with penetration depth.} compared to CR protons will diminish the photon flux relative to that of secondary electrons within interstellar ices. 

Secondary electrons are also produced within interstellar ices by other interstellar ionizing radiation such as $\gamma$ rays, X-rays, and vacuum UV photons
as well as by other components of Galactic CRs such as alpha particles and electrons.
In addition, in a region similar to that in which our sun may have been born, the CRIR has been found to be three orders of magnitude higher than the typical interstellar value \cite{Ceccarelli_2014,fontani_seeds_2017}. Because of the reasons enumerated above, \textit{we take our estimated secondary electron flux as a lower bound to the ionizing radiation-induced secondary electron flux within interstellar ices}. Given that electron-induced dissociation processes typically have higher cross-sections than those of photons, our order-of-magnitude secondary electron flux calculations suggest that the effects of ionizing-radiation-induced low-energy secondary electrons are at least as significant as those of photons in the interstellar synthesis of prebiotic molecules whose delivery by comets and meteorites likely kick-started life on earth. 

Although our calculations are only directly related to sub-micron-size interstellar ice particles, our results also have implications for ices in extrasolar and solar planets (e.g., Mars), dwarf planets (e.g., Pluto), moons (e.g., Europa), asteroids (e.g., Ceres), and comets (e.g., 67P/Churyumov–Gerasimenko). For example, given the orders of magnitude higher penetration depth of Galactic CRs compared to UV photons on the Martian surface, low-energy secondary electrons likely dominate in the subsurface Martian radiolysis, which has recently been hypothesized as a possible source of metabolic energy \cite{atri_investigating_2020}. Galactic CRs penetrate tens of meters below cometary surfaces facilitating radiolysis of cometary nuclei \cite{gronoff_effect_2020,maggiolo_effect_2020}. Our calculations indicate that low-energy electrons should dominate chemical modifications in such environs. 

In our planned future work, we will undertake a more detailed analysis that takes into account that energy deposition by high-energy charged particles is a random process that consists of numerous events, with each event transferring a small amount of energy. Our ongoing calculations involve Geant4-DNA (GEometry ANd Tracking 4-DNA) which is an extension of the Geant4 toolkit \cite{ALLISON2016186}, to model the passage of particles through matter. These calculations (to be published) will provide us with the total number of low-energy secondary electrons produced by high-energy charged particles as well as the energy distribution of the secondary electrons. Moreover, the Geant4 calculations will yield microscopic information (e.g., location of ionization) that could be employed to model chemical reactions that follow energy deposition. Nevertheless, given the three orders of magnitude variations in the measured CRIR that determines the ionizing radiation flux incident on interstellar ices, the order-of-magnitude calculations in this work are sufficient to demonstrate the importance of low-energy secondary electrons vis-\`{a}-vis non-ionizing photons in the extraterrestrial synthesis of complex organic molecules.

\section{Conclusions}
Because of the primacy of low-energy secondary electrons in all radiation chemistry processes, we have studied the potential role of low-energy electrons in astrochemistry. We estimate the flux of Galactic CR-induced secondary electrons in interstellar ices within dark, dense molecular clouds by considering (1) the CR spectra that best reproduce the CRIR in diffuse interstellar clouds, (2) the attenuated CR particle spectra after propagation through dark, dense molecular clouds, and (3) data from the NIST databases which provide information on the energy loss for charged particles traversing water. The results based on the attenuated Galactic CR spectrum indicate that the flux of low-energy electrons within these interstellar ices is almost as substantial as the flux of UV non-ionizing photons incident on ices in dark, dense molecular clouds. In some star-forming regions where the CRIR due to internal high-energy charged particles has been found to be up to three orders of magnitude higher than the typical interstellar values, low-energy secondary electron flux will dominate that of non-ionizing photons. Given that low-energy electrons likely have larger reaction cross-sections than photons, we argue that astrochemical models should consider the role of low-energy electrons ($<$ 20 eV) in energetic ice processing leading to the extraterrestrial synthesis of complex organic molecules. This information is crucial for understanding the processes involved in forming interstellar prebiotic molecules, which may have played a role in the emergence of life.

\section{Acknowledgements}
We thank Amanda Cheng for assistance with figure preparation. J.B.R.B's and C.R.A.’s work was supported by a grant from the National Science Foundation (NSF grant number CHE-1955215). C.R.A's work was also supported by Wellesley College (Faculty Awards and Brachman Hoffman small grants). Additionally, this research is partially funded by the Gordon and Betty Moore Foundation through Grant GBMF11565 and grant DOI https://doi.org/10.37807/GBMF11565.
The Massachusetts Space Grant Consortium supported the work of H.A. and A.K.W. K.B. gratefully acknowledges funding from the Arnold and Mabel Beckman Foundation.

\bibliography{cr.bib}
\end{document}